\documentclass[8.5pt,twoside,twocolumn]{article}
\usepackage[super,sort&compress,comma]{natbib}
\usepackage{mhchem}
\usepackage{times,mathptmx}
\usepackage{sectsty}
\usepackage{balance}

\usepackage{graphicx} %eps figures can be used instead
\usepackage{lastpage}
\usepackage[format=plain,justification=raggedright,singlelinecheck=false,font=small,labelfont=bf,labelsep=space]{caption} 
\usepackage{fancyhdr}
\begin{document}

\thispagestyle{plain}
\fancypagestyle{plain}{
\fancyhead[L]{\includegraphics[height=8pt]{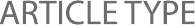}}
\fancyhead[C]{\hspace{-1cm}\includegraphics[height=20pt]{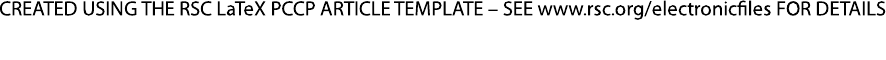}}
\fancyhead[R]{\includegraphics[height=10pt]{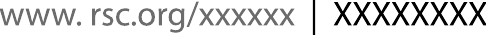}\vspace{-0.2cm}}
\renewcommand{\headrulewidth}{1pt}}
\renewcommand{\thefootnote}{\fnsymbol{footnote}}
\renewcommand\footnoterule{\vspace*{1pt}% 
\hrule width 3.4in height 0.4pt \vspace*{5pt}} 
\setcounter{secnumdepth}{5}

\makeatletter 
\def\subsubsection{\@startsection{subsubsection}{3}{10pt}{-1.25ex plus -1ex minus -.1ex}{0ex plus 0ex}{\normalsize\bf}} 
\def\paragraph{\@startsection{paragraph}{4}{10pt}{-1.25ex plus -1ex minus -.1ex}{0ex plus 0ex}{\normalsize\textit}} 
\renewcommand\@biblabel[1]{#1}            
\renewcommand\@makefntext[1]% 
{\noindent\makebox[0pt][r]{\@thefnmark\,}#1}
\makeatother 
\renewcommand{\figurename}{\small{Fig.}~}
\sectionfont{\large}
\subsectionfont{\normalsize} 

\fancyfoot{}
\fancyfoot[LO,RE]{\vspace{-7pt}\includegraphics[height=9pt]{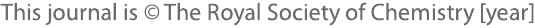}}
\fancyfoot[CO]{\vspace{-7.2pt}\hspace{12.2cm}\includegraphics{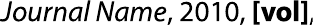}}
\fancyfoot[CE]{\vspace{-7.5pt}\hspace{-13.5cm}\includegraphics{headers/RF.pdf}}
\fancyfoot[RO]{\footnotesize{\sffamily{1--\pageref{LastPage} ~\textbar  \hspace{2pt}\thepage}}}
\fancyfoot[LE]{\footnotesize{\sffamily{\thepage~\textbar\hspace{3.45cm} 1--\pageref{LastPage}}}}
\fancyhead{}
\renewcommand{\headrulewidth}{1pt} 
\renewcommand{\footrulewidth}{1pt}
\setlength{\arrayrulewidth}{1pt}
\setlength{\columnsep}{6.5mm}
\setlength\bibsep{1pt}

\twocolumn[
  \begin{@twocolumnfalse}
\noindent\LARGE{\textbf{Phase behaviour of active Brownian particles: The role of dimensionality$^\dag$}}
\vspace{0.6cm}

\noindent\large{\textbf{Joakim Stenhammar,$^{\ast}$\textit{$^{a,b}$} Davide Marenduzzo,\textit{$^{a}$} Rosalind J. Allen,\textit{$^{a}$} and Michael E. Cates\textit{$^{a}$}}}\vspace{0.5cm}

%Please note that \ast indicates the corresponding author(s) but no footnote text is required. 

\noindent\textit{\small{\textbf{Received Xth XXXXXXXXXX 20XX, Accepted Xth XXXXXXXXX 20XX\newline
First published on the web Xth XXXXXXXXXX 200X}}}

\noindent \textbf{\small{DOI: 10.1039/b000000x}}
\vspace{0.6cm}

\noindent \normalsize{Recently, there has been much interest in activity-induced phase separations in concentrated suspensions of ``active Brownian particles'' (ABPs), self-propelled spherical particles whose direction of motion relaxes through thermal rotational diffusion. To date, almost all these studies have been restricted to 2 dimensions. In this work we study activity-induced phase separation in 3D and compare the results with previous and new 2D simulations. To this end, we performed state-of-the-art Brownian dynamics simulations of up to 40 million ABPs -- such very large system sizes are unavoidable to evade finite size effects in 3D. Our results confirm the picture established for 2D systems in which an activity-induced phase separation occurs, with strong analogies to equilibrium gas-liquid spinodal decomposition, in spite of the purely non-equilibrium nature of the driving force behind the phase separation. However, we also find important differences between the 2D and 3D cases. Firstly, the shape and position of the phase boundaries is markedly different for the two cases. Secondly, for the 3D coarsening kinetics we find that the domain size grows in time according to the classical diffusive $t^{1/3}$ law, in contrast to the nonstandard subdiffusive exponent observed in 2D.}

\vspace{0.5cm}
 \end{@twocolumnfalse}
  ]

%Footnotes
\footnotetext{\dag~Electronic Supplementary Information (ESI) available: Movies showing the time evolution of the density field obtained from Brownian dynamics simulations of ABPs and solutions of the continuum model in 2D and 3D. See DOI: 10.1039/b000000x/}

%Please use \dag to cite the ESI in the main text of the article.
%If you article does not have ESI please remove the the \dag symbol from the title and the above footnotetext.

\footnotetext{\textit{$^{a}$~SUPA, School of Physics and Astronomy, University of Edinburgh, JCMB Kings Buildings, Edinburgh EH9 3JZ, United Kingdom; E-mail: j.stenhammar@ed.ac.uk}}
\footnotetext{\textit{$^{b}$~Division of Physical Chemistry, Department of Chemistry, Lund University, P.O. Box 124, S-221 00 Lund, Sweden}}

\section{Introduction}
Active materials, whose constituents possess the ability to convert chemical energy into work, have recently been shown to exhibit a range of exotic behaviours compared to what is observed in passive systems.\cite{Marchetti-2013,Romanczuk-2012,Cates-2012} These include rectification of bacterial motion, \cite{Galajda-2007,Angelani-2009,Lambert-2010,Potosky-2013} giant density fluctuations, \cite{Narayan-2007,Deseigne-2010,Zhang-2010,Wensink-2012} dynamic phase transitions, \cite{Cates-2010,Theurkauff-2012,McCandlish-2012,Farrell-2012,Redner-2013-PRE,Buttinoni-2013} and the ability to power microscopic motors.\cite{Angelani-2009,DiLeonardo-2010} From the perspective of statistical thermodynamics, all these phenomena are made possible by the fact that active matter systems are intrinsically far from equilibrium even at steady state, and thus are not constrained by the strict rules imposed on systems that obey detailed balance. 

The classic example of active particles is that of ``run-and-tumble'' bacteria, such as \emph{Escherichia coli}, whose ballistic motion (``runs'') is regularly punctuated by random reorientations of the swimming direction (``tumbles''), occurring with a frequency $\alpha$. On time-scales much larger than $\alpha^{-1}$, these dynamics lead to a persistent random walk with a characteristic step length $v_0/\alpha$, where $v_0$ is the swim speed of a single bacterium. Because of this, the long-time behaviour of a run-and-tumble bacterium is similar to that of a diffusing Brownian particle, with an effective diffusion coefficient $D$ given by
\begin{equation}\label{D_0_alpha}
D = \frac{v_0^{2}}{d\alpha},
\end{equation}
where $d$ is the spatial dimensionality of the system; for realistic parameter values, $D$ is hundreds of times larger than the Brownian diffusivity of a similarly sized colloidal particle.\cite{Cates-2012}

A second important class of active particles is that of active Brownian particles (ABPs), whose swimming direction continuously relaxes through thermal rotational diffusion rather than through discrete tumbling events. The prime experimental example of such ABPs is that of ``catalytic swimmers'', half-coated spherical colloids rendered motile through a surface reaction.\citep{Howse-2007,Ebbens-2010} As shown by Cates and Tailleur, \cite{Cates-2013} the long-time dynamics of ABPs are equivalent to those of run-and-tumble particles upon the substitution $\alpha \leftrightarrow (d-1)D_{\mathrm{r}}$, where $D_{\mathrm{r}}$ is the (thermal) rotational diffusion constant.\footnote{Note that ABPs, unlike run-and-tumble particles, cannot be realized in 1D.} 

Due to the fact that active particles exhibit diffusive behaviour at long length- and timescales, mass transport in such suspensions is governed by laws similar to those operating in equilibrium Brownian systems. In particular, in a suspension of non-interacting self-propelled particles with spatially uniform swim speeds and tumble rates, transport is governed by the standard form of Fick's first law for ideal systems, relating the mass flux $\mathbf{J}$ to the (effective) diffusivity $D$ and concentration gradient $\nabla \rho$:
\begin{equation}\label{Fick_eq}
\mathbf{J} = -D \nabla \rho.
\end{equation}
For a spatially uniform diffusivity $D$, Eq. \eqref{Fick_eq} ensures relaxation towards a state of homogeneous density ($\nabla \rho = 0$). However, as shown by Schnitzer, \cite{Schnitzer-1993} if the effective diffusivity of Eq. \eqref{D_0_alpha} varies in space through a non-uniform swim speed (\emph{e.g.} due to an inhomogeneous fuel concentration), the situation becomes markedly different. This position-dependent swimming velocity $v(\mathbf{r})$ leads to the following generalization of Eq. \eqref{Fick_eq}:  
\begin{equation}\label{Fick_RTP}
\mathbf{J} = -D(\mathbf{r}) \nabla \rho - \frac{\rho v(\mathbf{r})}{d\alpha} \nabla v.
\end{equation}

Tailleur and Cates \citep{Tailleur-2008,Cates-2013} considered an analogue of Eq. \eqref{Fick_RTP} for the case when particles instead slow down due to collisions and steric interactions, leading to a swim-speed that depends on the local density of bacteria, \emph{i.e.} $v(\mathbf{r}) = v(\rho(\mathbf{r}))$. This assumption leads to the following form of the flux:
\begin{equation}\label{Fick_v_rho}
\mathbf{J} = -\left[ D(\rho) + \frac{\rho v(\rho)}{d\alpha} \frac{\mathrm{d}v}{\mathrm{d}\rho} \right] \nabla \rho.
\end{equation}
For swim speeds that decrease steeply enough with density, the term in square brackets in Eq. \eqref{Fick_v_rho} can become negative, enabling the emergence of non-uniform steady states. Microscopically, this corresponds to an activity-induced phase separation induced by a kinetic feedback mechanism, whereby a local (positive) density fluctuation will lead to a local slowdown of particles, which in turn causes a further accumulation of particles, eventually leading to the nucleation of a dense phase. The existence of this motility-induced phase transition has recently been confirmed for purely repulsive ABPs in both experiments \cite{Buttinoni-2013} and simulations. \cite{Fily-2012,Redner-2013} Furthermore, Eq. \eqref{Fick_v_rho} together with a carefully tuned noise term and a specific form of $v(\rho)$, was recently shown to yield predictions in quantitative agreement with the structural and kinetic properties of a phase-separating ABP fluid.\cite{Stenhammar-2013} 

The vast majority of previous work on activity-induced phase separation (with a single, very recent, exception~\citep{Wysocki-2013}) has focused on phenomena in 2 dimensions. However, real suspensions of active particles are often 3-dimensional. Our goal here is therefore to investigate in detail phase separation in concentrated 3D suspensions of ABPs. From a theoretical standpoint, a phase separating ABP suspension can be viewed as a binary fluid, where the two components form the dilute and dense phase. A key difference in the topology of binary mixtures in 2D and in 3D is that in 2D fluid bicontinuity is only possible by fine tuning to a single composition (50:50 if the fluid is otherwise symmetric), meaning that the generic situation in 2D is that of disconnected droplets of the minority phase inside a matrix of the majority phase.\cite{Wagner-1998,Wagner-1999} In 3D, however, both fluids can remain continuously connected through the whole sample for a wide range of compositions. In general, the interplay of such dimensionality-dependent topological differences with nonequilibrium physics is subtle and difficult to predict \emph{a priori};\cite{Stansell-2006,Stratford-2007} the specific case of active systems is similarly difficult to gauge. For instance, self-propelled particles might move quite differently in disconnected dilute pockets, as would be expected in a concentrated 2D suspension, compared to in a percolating dilute phase as may be expected in 3D at a similar density. Another important aspect is that of fluctuations: on general grounds, the role of noise is expected to decrease as the dimensionality increases, and previous work in 2D reported important effects of fluctuations on the phenomenology of phase separation.~\cite{Stenhammar-2013}

In the present study, we perform massively parallel Brownian dynamics simulations of up to $\sim 4 \times 10^{7}$ repulsive ABPs using $\sim 8000$ CPUs. Such large systems are required to avoid finite size effects which would otherwise compromise our conclusions. We then compare our results to those obtained from large-scale 2D ABP simulations, similar to those performed previously.\citep{Fily-2012,Redner-2013,Stenhammar-2013} Our simulation results are furthermore compared to results obtained from numerically solving a recently developed continuum model,\cite{Stenhammar-2013} which we first extend from 2D to 3D. The results obtained from both particle and continuum simulations confirm that activity-induced phase separations have remarkable analogies to classical liquid-gas transitions, despite their completely different physical origins. Furthermore, we find these analogies to be even more marked in 3D than in 2D, as indicated for instance by the kinetics of domain growth, which in 3D follows the classical $t^{1/3}$ growth law~\cite{Chaikin} from relatively early times. This is, we hypothesise, because fluctuations are less important in this 3D scenario. We also find other important differences between the physics of the 2D and 3D systems: for instance, the shape of the phase diagram and the region within which phase separation is observed are significantly different. As the phase separation occurs, the morphologies of the growing domains are also qualitatively different, with the 3D case featuring smoother interfaces and the near-absence of disconnected droplet phases. 

\section{Discrete ABP model}
Our microscopic ABP model consists of spherical particles interacting through a repulsive, pairwise additive Weeks-Chandler-Andersen potential, given by 
\begin{equation}\label{WCA}
U = 4\varepsilon \left[ \left(\frac{\sigma}{r}\right)^{12} - \left(\frac{\sigma}{r}\right)^{6} \right] + \varepsilon 
\end{equation}
with an upper cut-off at $r = 2^{1/6}\sigma$, beyond which $U = 0$. Here $\sigma$ denotes the  particle diameter, $\varepsilon$ determines the interaction strength, and $r$ is the center-to-center separation between two particles. The model was studied by solving the fully overdamped translational and rotational Langevin equations:
\begin{align}
\partial_t \mathbf{r}_i &= \beta D_{\mathrm{t}} \left[ \mathbf{F}_i + F_{\mathrm{p}} \mathbf{p}_i \right] + \sqrt{2D_{\mathrm{t}}}{\boldsymbol \Lambda_{\mathbf{r}}} \label{Langevin_t} \\
\partial_t \theta_i &= \sqrt{2D_{\mathrm{r}}} \Lambda_{\theta} & \textrm{(2D)} \label{Langevin_r_2D}  \\
\partial_t \mathbf{p}_i &= \sqrt{2D_{\mathrm{r}}} (\mathbf{p}_i \times {\boldsymbol \Lambda_{\mathbf{p}}}) & \textrm{(3D)} \label{Langevin_r_3D}
\end{align}
where $\mathbf{F}_i$ is the total conservative force acting on particle $i$, $F_{\mathrm{p}}$ is the (constant) magnitude of the self-propulsion force, whose direction is defined by $\mathbf{p}_i$ [where $\mathbf{p}_i = (\cos \theta_i, \sin \theta_i)$ in 2D], $D_{\mathrm{t}}$ and $D_{\mathrm{r}} = 3D_{\mathrm{t}}/\sigma^{2}$ denote the translational and rotational diffusivities, $\beta = (k_{\mathrm{B}} T)^{-1}$ is the inverse thermal energy, and ${\boldsymbol \Lambda_{\mathbf{r}}}$, $\Lambda_{\theta}$, and ${\boldsymbol\Lambda_{\mathbf{p}}}$ are unit-variance stochastic vectors of appropriate dimensionality, whose Cartesian components $\Lambda_i$ satisfy $\langle \Lambda_i(\mathbf{r},t) \Lambda_j(\mathbf{r}',t') \rangle = \delta_{ij}\delta(\mathbf{r}-\mathbf{r}') \delta(t-t')$. All simulations were carried out using the LAMMPS \cite{Plimpton-1995} molecular dynamics software package, with system sizes of up to $1000 \sigma$ ($N \approx 7 \times 10^{5}$) and $350 \sigma$ ($N \approx 4 \times 10^{7}$) in 2D and 3D, respectively. Further simulation details are presented in Section \ref{ABP_section}. 

\section{Continuum model}\label{Continuum_derivation}

We begin by reviewing the main aspects of the continuum theories presented in Refs. \cite{Tailleur-2008} and, \cite{Stenhammar-2013} generalising to spatial dimensionalities $d > 2$.

We begin by noting that Eq. \eqref{Fick_v_rho} can be reexpressed in an equilibrium-like form as follows:
\begin{equation}\label{Flux_F}
\mathbf{J} = - \rho D(\rho) \nabla \left[ \frac{\delta \mathcal{F}_0}{\delta \rho} \right],
\end{equation}
where $\mathcal{F}_0 = \int f_0 \mathrm{d}\mathbf{r}$ is an effective free energy, and 
\begin{equation}\label{f_0}
f_0 = \rho (\ln \rho - 1) + \int_{0}^{\rho} \ln [v(u)] \mathrm{d}u
\end{equation}
defines the corresponding free-energy density, composed of an ideal entropy-like term and a $v$-dependent term that resembles an enthalpic attraction. Eqs. \eqref{Flux_F} -- \eqref{f_0} clearly show how a density-dependent swim speed can induce phase separation into high- and low-density phases, and suggest that the form of this transition may show analogies with equilibrium phase transitions, in spite of the strongly non-equilibrium nature of the self-trapping mechanism leading to the instability. 

As previously derived for 1D run-and-tumble particles\cite{Tailleur-2008} and generalized to ABPs in higher dimensions,\cite{Cates-2013} the coarse-grained density field $\rho$ of particles with a density-dependent swim speed obeys:
\begin{equation}\label{rho_t}
\partial_t \rho = -\nabla \cdot \mathbf{J} = -\nabla\cdot \left\{ -D(\rho) \rho \nabla \mu + \sqrt{2 D(\rho) \rho} {\boldsymbol \Lambda} \right\}.
\end{equation}
Here $D(\rho)$ is an effective one-body diffusivity, $\mu$ an effective chemical potential, and ${\boldsymbol \Lambda}$ is a 3-dimensional random vector as discussed following Eq. \eqref{Langevin_r_3D}. The effective bulk chemical potential $\mu_0$ is given by the derivative of Eq. \eqref{f_0}:
\begin{equation}\label{mu_0}
\mu_0(\rho) \equiv \frac{\delta \mathcal{F}_0}{\delta \rho} = \ln\rho + \ln v(\rho).
\end{equation} 

To enable a full analysis of phase-separation kinetics in the same spirit as the classical Cahn-Hilliard equation,\cite{Chaikin,Puri} Eq. \eqref{mu_0} needs to be complemented with an interfacial energy-like term that stabilizes domain walls between the phases. Following Ref. \cite{Stenhammar-2013} we accomplish this by assuming that a single ABP samples the local density over a length scale proportional to the density-dependent persistence length $\ell(\rho)$ of ABP trajectories, given by
\begin{equation}
\ell(\rho) =  \frac{\tau_{\mathrm{r}} v(\rho)}{d-1},
\end{equation}
where $\tau_{\mathrm{r}} = D^{-1}_{\mathrm{r}}$ is the rotational relaxation time. As we previously showed in Ref. \cite{Stenhammar-2013},  this leads to an additional term in the effective chemical potential:
\begin{equation}\label{mu_kappa}
\mu = \mu_0 - \kappa(\rho)\nabla^2\rho,
\end{equation}
where
\begin{equation}\label{kappa}
\kappa(\rho) = - \left[\frac{\gamma_{0} \tau_{\mathrm{r}}}{d-1}\right]^{2} v(\rho) \frac{\mathrm{d} v}{\mathrm{d} \rho},
\end{equation}
with $\gamma_0$ a dimensionless order-unity free parameter. Unlike a traditional density-independent interfacial energy, the second term of Eq. $\eqref{mu_kappa}$ cannot be written as the derivative of a free-energy functional; thus, the effective chemical potential in Eq. \eqref{mu_kappa} violates detailed balance at second order in a gradient expansion. However, detailed balance can be restored through the modification 
\begin{equation}\label{mu_db}
\mu_{\mathrm{DB}} = \mu_0 - \kappa(\rho)\nabla^2\rho - \frac{\mathrm{d}\kappa}{\mathrm{d}\rho}\frac{(\nabla\rho)^2}{2},
\end{equation}
which again renders $\mu$ integrable; this allows comparison between models that differ \emph{only} in whether detailed balance applies or not. 

In order to emulate the physics of excluded volume interactions between ABPs, and thus to prevent the emergence of a phase of infinite density, a repulsive term has to be added to the effective bulk free energy, \emph{i.e.}, $f_0 \to f_0 + f_{\mathrm{rep}}$. As previously,\cite{Stenhammar-2013} we choose the following quartic form of the repulsive contribution $f_{\mathrm{rep}}$: 
\begin{equation}\label{f_rep}
f_{\mathrm{rep}} = k_{\mathrm{rep}} \Theta(\rho - \rho_{\mathrm{t}})(\rho-\rho_{\mathrm{t}})^{4}.
\end{equation}
In Eq. \eqref{f_rep}, $\Theta$ denotes the Heaviside step function, $k_{\mathrm{rep}}$ determines the strength of the repulsion, and $\rho_{\mathrm{t}}$ is the threshold density at which the repulsion is switched on. In practice, both these quantities are treated as free parameters, but they were previously found not to significantly affect the phase separation dynamics.\cite{Stenhammar-2013} 

The crucial component in the continuum model is the form of the density-dependent swim speed $v(\rho)$. To determine this, we make the assumption that single ABP trajectories consist of straight runs with speed $v_0 = v(0)$ interrupted by collision events of duration $\tau_{\mathrm{c}}$, each leading to a complete stall of the particle motion. At low and intermediate densities this assumption leads to the following form for $v(\rho)$: \cite{Stenhammar-2013}
\begin{equation}\label{v_rho}
v(\rho) = v_0 \left(1 - \frac{\tau_{\mathrm{c}}}{\tau_{\mathrm{MF}}}\right) = v_0 (1 - v_0 \sigma_{\mathrm{s}} \tau_{\mathrm{c}} \rho  ),
\end{equation}
where $\sigma_{\mathrm{s}}$ is a scattering cross-section and $\tau_{\mathrm{MF}}$ is the mean free time between two collisions. Furthermore, the effective single-particle diffusivity $D(\rho)$ follows as [\emph{cf}. Eq. \eqref{D_0_alpha}]
\begin{equation}\label{D_eff}
D(\rho) = \frac{v^{2}(\rho)\tau_{\mathrm{r}}}{d(d-1)} = D_0(1-v_0 \sigma_{\mathrm{s}} \tau_{\mathrm{c}}\rho)^{2},
\end{equation}
where $D_0 = D(0)= v_0^{2} \tau_\mathrm{r} [d(d-1)]^{-1}$. The same functional form (\emph{i.e.}, a linear decrease of $v$ with density) has recently been observed\cite{Fily-2012} and theoretically predicted\cite{Bialke-2013,Fily-2013} elsewhere. As described in Appendix \ref{v_rho_section}, $v(\rho)$ and $D(\rho)$ can be accurately measured in discrete ABP simulations performed in the one-phase region of the phase diagram and fitted to the functions $v(\phi) = v_0 (1-a\phi)$ and $D(\phi) = D_0 (1-b\phi)^{2}$, where $\phi \in [0,1]$ is the particle packing fraction and $a \simeq b$ are fitting parameters.\footnote{From the kinetic argument above, $a$ and $b$ are predicted to be identical, which is also what we found in our previous 2D study where $a = 1.05$ and $b = 1.04$. \cite{Stenhammar-2013} In the rest of the article the two parameters will therefore be assumed identical and both denoted by $a$.}

Non-dimensionalizing Eqs. \eqref{rho_t} -- \eqref{mu_kappa} in terms of the length and time units $D_0 / v_0 \equiv \lambda$ and $D_0 / v^{2}_0 = \tau_r / d(d-1)$ (equivalent to fixing $D_0 = v_0 = 1$), respectively, finally gives
\begin{equation}\label{rho_t_rescaled}
\partial_t \phi = \nabla\cdot \left\{\phi (1-a\phi)^{2} \nabla \mu- \sqrt{2 \phi (1-a\phi)^{2} N_0^{-1}} {\boldsymbol \Lambda} \right\}
\end{equation}
\begin{align}
\mu_0 &= \ln \left[\phi(1-a\phi) \right] \label{mu_0_final} \\
\mu_{\mathrm{rep}} &= 4k_{\mathrm{rep}}\Theta(\phi - \phi_{\mathrm{t}})(\phi - \phi_{\mathrm{t}})^{3} \\
\mu &= \mu_0 +\mu_{\mathrm{rep}}-\kappa_0(1-a\phi)\nabla^{2}\phi, \label{mu_nabla_final}
\end{align}
where $\kappa_0 = a\left[(d-1)^{-1}\gamma_0\tau_{\mathrm{r}}\right]^{2}$. Since the order-unity factor $\gamma_0$ is unknown, $\kappa_0$ will be treated as a free parameter. Furthermore, $N_0 = \lambda^{d}/V_{\mathrm{p}}$ is the number of particles in a cube of side length $\lambda$ at nominal packing fraction unity, $V_{\mathrm{p}}$ being the volume of a single particle. Note that a mapping between the continuum model and discrete ABP simulations emerges automatically, including the strength of the noise, with the choice of units made here. The only free parameters in the model are thus $\kappa_0$, which controls the strength of the interfacial-like term, and $k_{\mathrm{rep}}$ and $\phi_{\mathrm{t}}$, that control the strength and onset of the repulsive free energy, respectively. 

\section{Results and discussion}

\subsection{Phase diagram}\label{Phase_diagram_section}

\begin{figure}[h!]
  \centering
  \includegraphics[height=4.1cm]{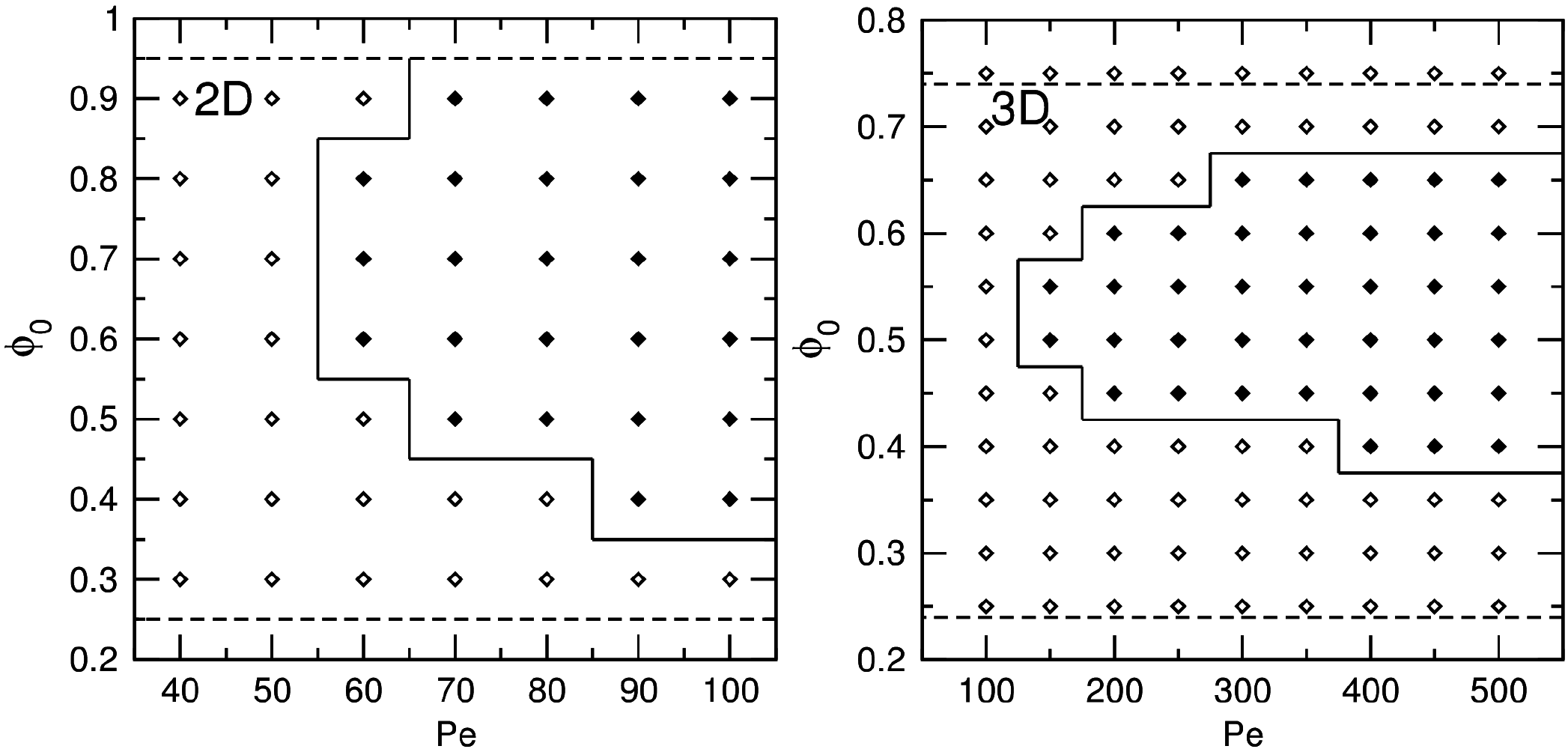}
  \caption{Phase diagrams in the $\mathrm{Pe} -- \phi_0$ plane as determined from ABP simulations of systems with size $L_{\mathrm{box}} = 150\sigma$ (2D, left panel) and $L_{\mathrm{box}} = 30\sigma$ (3D, right panel). Open symbols denote state points with a homogeneous density, and filled symbols denote points with phase-separated steady states. Solid lines show the approximate spinodals, while the dashed lines denote the approximate high-Pe asymptotes for the binodal densities, determined for Pe $ = 100$ (2D) and Pe $ = 300$ (3D). The locations of the spinodals were determined both by visual inspection and by examining the growth in the characteristic length-scale $L(t)$, as further described in Section \ref{Kinetics_section}. The lower binodal densities (0.25 and 0.24 in 2 and 3 dimensions, respectively) were determined using larger systems ($L_{\mathrm{box}} = 1000\sigma$ and $L_{\mathrm{box}} = 100\sigma$ in 2D and 3D, respectively) by starting from fully phase-separated states, while the upper boundaries were determined by approximate extrapolation to the case of vanishing volume of the dilute phase.}
 \label{phase_diagrams}
\end{figure}

Fig. \ref{phase_diagrams} shows 2D and 3D phase diagrams as a function of the average particle packing fraction $\phi_0$ and the P\'eclet number, Pe, defined by
\begin{equation}
\mathrm{Pe} \equiv \frac{3v_0 \tau_{\mathrm{r}}}{\sigma},
\end{equation}
where $\sigma$ is the particle diameter. Clearly, Pe plays a role similar to that of the inverse temperature in equilibrium phase diagrams, in accordance with what has been observed previously in 2D.\cite{Redner-2013,Stenhammar-2013} The computational cost of accurately predicting phase diagrams for phase separating fluids is very large; hence, the system sizes used for determining these are, particularly in 3D, smaller than what is needed to guarantee the absence of finite-size effects. Thus, Fig.~\ref{phase_diagrams} should be viewed as providing approximate phase maps. Moreover, as will be further discussed in the next subsection, our continuum theory is not capable of accurately reproducing the binodals and/or spinodals of this system; thus, no quantitative comparison between ABP and continuum theory will be attempted regarding the phase diagrams (as opposed to the phase separation kinetics).

The 2D phase diagram broadly resembles that determined previously\cite{Redner-2013} for an essentially identical ABP model, while the 3D phase diagram is in qualitative agreement with that determined recently\cite{Wysocki-2013} using a Yukawa-type pair potential to model the ABPs. Perhaps most strikingly, it is clear that significantly larger P\'eclet numbers are needed for phase separation to occur in 3D compared to in 2D. In experiments, where the value of Pe is limited and difficult to vary (for a bacterial strain it is fixed by the swim speed: for \emph{E. coli}, Pe would be at most $\sim 100$), and where concentrated suspensions are not easy to achieve, this fact may well provide practical obstacles to the observation of phase separation in 3D. Qualitatively, the difference in the critical Pe for phase separation between 2D and 3D may be explained by the fact that orientational correlations decay faster in 3D than in 2D for a given value of $\tau_{\mathrm{r}}$; thus, a higher P\'eclet number is needed to accomplish a given collision time $\tau_{\mathrm{c}}$, which is the parameter that determines the critical Pe for phase separation.\citep{Redner-2013} 

We also observe that there are both binodal and spinodal lines in both phase diagrams in Fig.~\ref{phase_diagrams}, \emph{i.e.}, there are regions between these lines in which the fluid will not spontaneously phase separate from a homogeneous suspension, but will remain phase separated when initialized from a two-phase configuration. This metastable region is furthermore slightly larger in 3D than in 2D. Note, however, that the calculated ``binodals'' are actually asymptotes representing the high-Pe limit, and were obtained from simulations with $\mathrm{Pe} = 100$ and $\mathrm{Pe} = 300$ in 2D and 3D, respectively, by investigating the stability of the phase-separated state when starting from an initial configuration consisting of a single close-packed domain. Thus, we do not assess the behaviour of this boundary at low Pe, including whether the binodals and spinodals will merge at the critical point as is the case in equilibrium phase diagrams. Nevertheless, the existence of both binodal and spinodal lines in the phase diagrams is once again in analogy with equilibrium phase diagrams, in spite of the purely non-equilibrium nature of the phase separation studied here. The relatively large size of the metastable region (compared to the size of the binodal region) is slightly surprising considering the fact that the apparent noise level (or, equivalently, ``effective temperature'') is significantly higher in the ABP systems studied here than in the corresponding thermal systems, something which should help overcoming any (effective) activation energies keeping the system in a metastable state. 

\begin{figure}[h!]
  \centering
  \includegraphics[height=5.5cm]{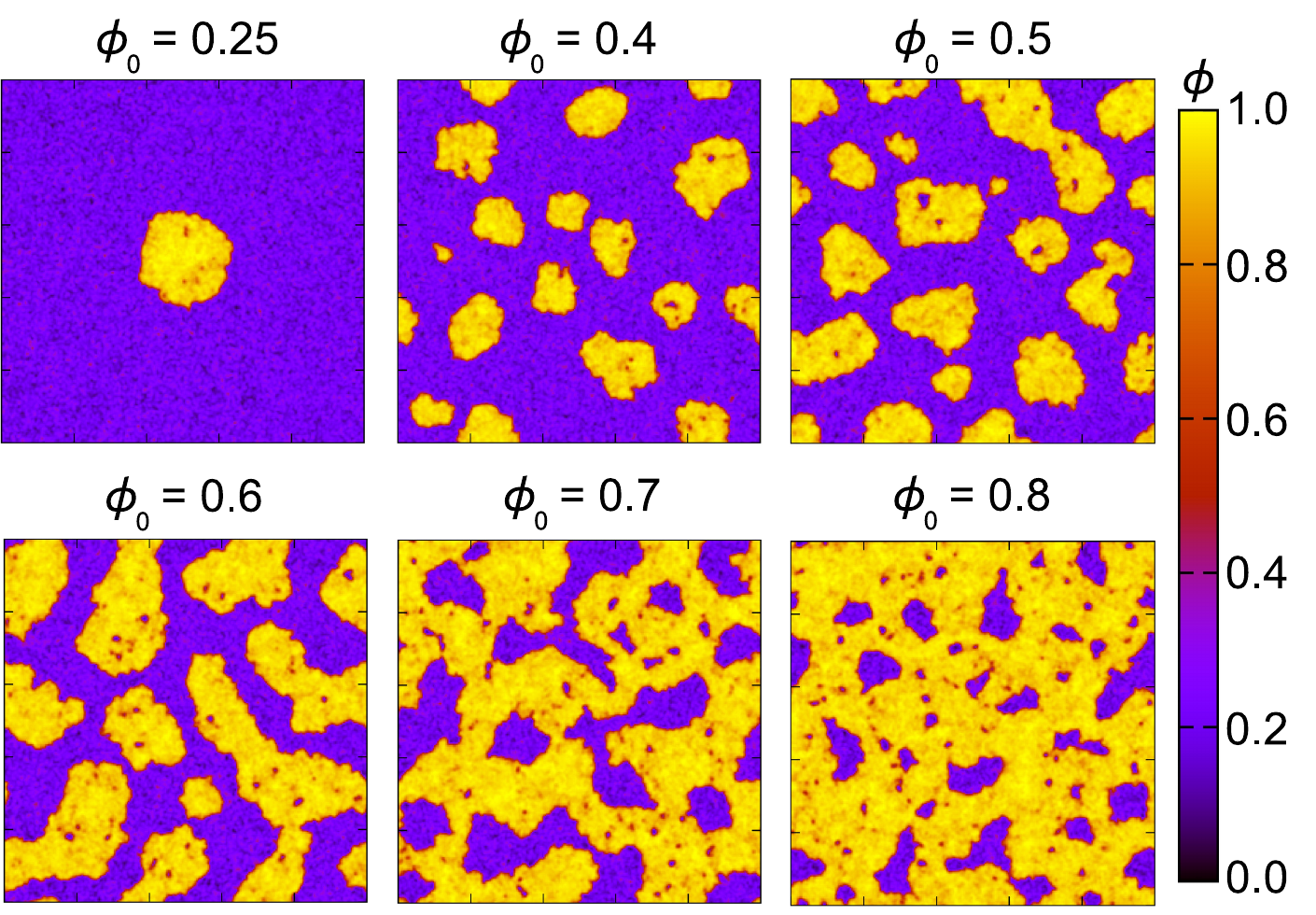}
  \caption{Snapshots taken at $t = 1000 \tau_{\mathrm{r}}$ obtained from 2D ABP simulations with varying overall area fraction $\phi_0$ as indicated. The systems are of size $L_{\mathrm{box}} = 1000 \sigma$ ($N \approx 7 \times 10^{5}$). The simulation with $\phi_0 = 0.25$ was started from a fully phase-separated state with a single close-packed circular domain, while the remaining runs were started from equilibrated (homogeneous) suspensions of passive particles. The density field was obtained by numerical coarse-graining on a grid, as described in Section \ref{ABP_section}.}
  \label{tieline_2D}
\end{figure}

\begin{figure}[h!]
  \centering
  \includegraphics[height=5.5cm]{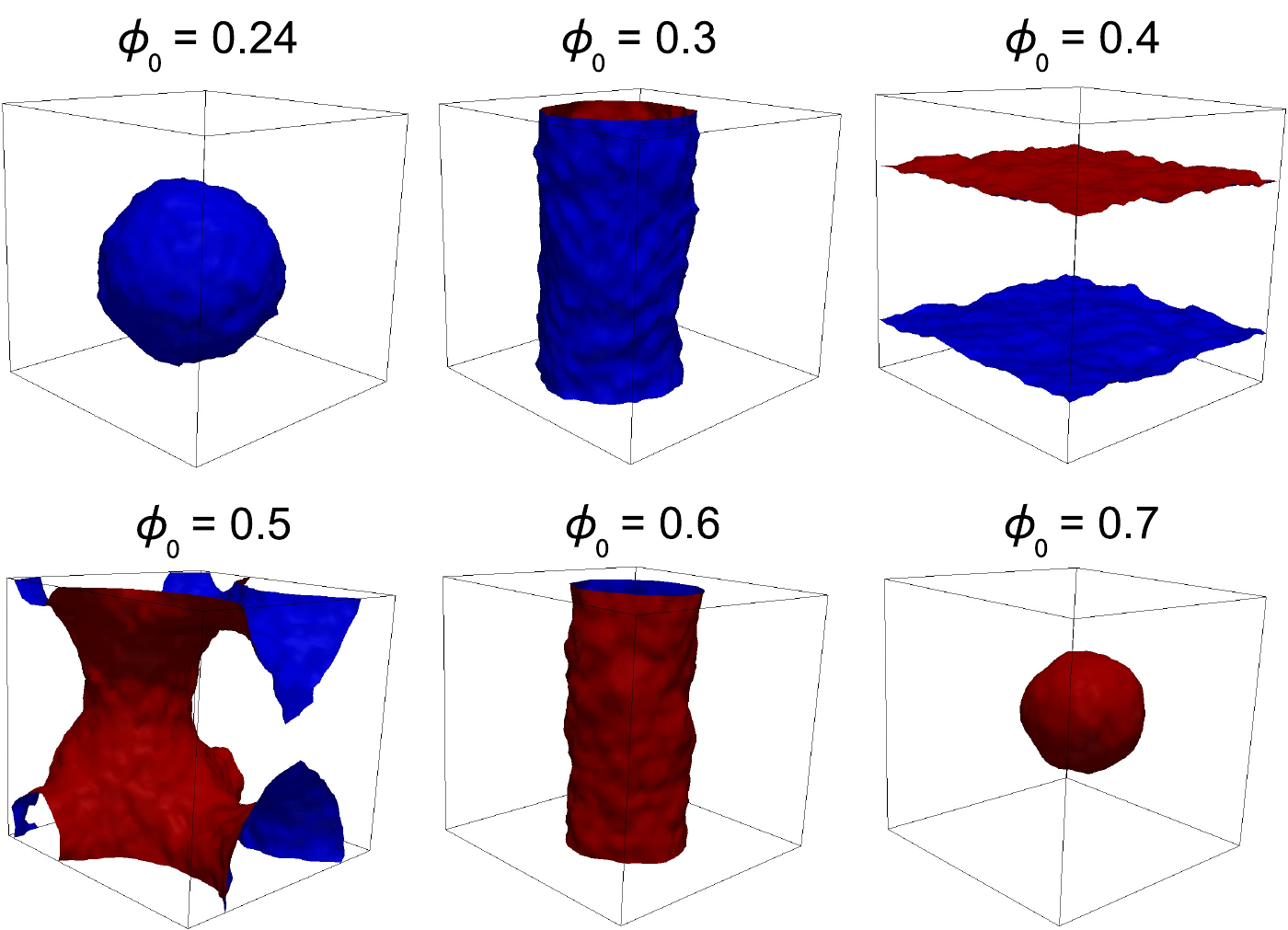}
  \caption{Snapshots taken at $t = 120 \tau_{\mathrm{r}}$ obtained from 3D ABP simulations with varying average volume fraction $\phi_0$ as indicated. The systems are of size $L_{\mathrm{box}} = 100 \sigma$ ($N \approx 10^{6}$). The interfaces represent isosurfaces drawn at $\phi = 0.4$, where the red side faces the dense phase and the blue side faces the dilute phase. The simulations with $\phi_0 = 0.24$, 0.3, and 0.7 were started from fully phase-separated states consisting of a single close-packed domain with a spherical interface for $\phi_0 = 0.24$ and with flat interfaces for $\phi_0 = 0.3$ and 0.7, while the remaining runs were started from equilibrated (homogeneous) suspensions of passive particles.}
  \label{tieline_3D}
\end{figure}

In Figs. \ref{tieline_2D} and \ref{tieline_3D}, representative snapshots taken along the density tieline of 2D and 3D phase-separating ABP systems well within the phase separated region are shown. These snapshots were obtained from much larger systems than those used to determine the phase diagrams, and were run for P\'eclet numbers of 100 and 300 in 2D and 3D, respectively. In 2D (Fig. \ref{tieline_2D}), the sequence of microstructures evolves from a phase of isolated dense droplets in a dilute background at low density, via an almost bicontinuous phase at intermediate packing fractions, to a phase exhibiting dilute droplets in a dense matrix at still higher densities. In 3D (Fig. \ref{tieline_3D}), the geometries of the phase separated structures are more well-defined, or, equivalently, the fluctuations at the interface between the dense and dilute phases are smaller. This effect is likely also connected with the apparently lower noise level in 3D compared to in 2D, which is clearly visible in the movies available as Electronic Supplementary Information$^\dag$. Furthermore, the evolving 3D morphologies always exhibit a single domain of each phase, unlike the corresponding 2D snapshots\footnote{However, our results indicate that the 2D systems will also eventually coarsen until a single domain of each phase remains. Thus, we do not see any sign of the finite cluster phases observed in experiments on active colloids. \citep{Theurkauff-2012,Palacci-2013}}. While the 3D structures seen in Fig. \ref{tieline_3D} are clearly influenced by the presence of periodic boundaries, they still convincingly suggest the appearance of spherical, cylindrical, flat, and saddle-like interfaces between the two phases. Furthermore, and as can be seen in Fig. \ref{tieline_2D} for the 2D case, the density of the two phases remains essentially constant along the tieline in both 2D and 3D,\footnote{Note, however, that the density $\phi_g$ of the dilute phase in similar systems has been predicted and observed\citep{Redner-2013,Abkenar-2013} to change with Pe as $\phi_g \sim \mathrm{Pe}^{-1}$.} an observation which is once again in accordance with the phenomenology of equilibrium phase transitions. 

A subtle point which needs to be highlighted in the present context is the way in which the P\'eclet number is varied. In most previous studies of ABPs (with one very recent exception\cite{Fily-2013} employing soft particles), Pe was varied by changing the bare swim speed $v_0$ while keeping the rotational relaxation time $\tau_{\mathrm{r}}$ constant. Viewing the problem from an energetic perspective, Pe quantifies the ratio between the ``ballistic energy'' $F_{\mathrm{p}} \sigma$ and the thermal energy $k_{\mathrm{B}}T$. For a system of infinitely hard spheres, these two are the only energy scales present, and thus Pe uniquely determines the balance between active and thermal forces. However, for the slightly softer spheres used here and in most previous ABP studies, another energy scale arises, describing the steepness of the repulsive potential; for the WCA potential used here, this energy scale is quantified through the Lennard-Jones parameter $\varepsilon$. Thus, Pe is no longer sufficient to describe the system, but the ratio $\varepsilon/k_{\mathrm{B}}T$, quantifying how ``hard'' the particles are, also comes into play. In previous studies on self-propelled WCA particles, $k_{\mathrm{B}}T / \varepsilon = 1$ was used throughout, and $F_{\mathrm{p}}$ (or, equivalently, $v_0$) was used as the free parameter controlling Pe. However, this implies that the ratio $F_{\mathrm{p}} \sigma / \varepsilon$ is not constant throughout the phase diagram, potentially leading to undesired effects. For example, an increase of $F_{\mathrm{p}} \sigma / \varepsilon$ will lead to a decreasing effective radius of the particles, meaning that colliding particles will exhibit increasingly large overlaps as Pe increases: for two ABPs with $\mathrm{Pe} = 300$ and $k_{\mathrm{B}}T / \varepsilon = 1$ colliding at a 90 degree angle, the center-to-center distance where the active and repulsive forces balance will be $\sim 0.85 \sigma$, \emph{i.e.}, 15 percent smaller than the ``thermal'' diameter $\sigma$. Apart from yielding an unphysical $\mathrm{Pe}$-dependent effective particle size, controlling Pe through this method will also lead to very large repulsive forces between overlapping particles. This effect may also help to explain the recently reported reentrant behaviour where the motility-induced phase transition is suppressed for large enough values of Pe.\cite{Bialke-2013} In order to avoid these effects, we chose to keep $F_{\mathrm{p}} \sigma / \varepsilon$ (and thus $v_0$) constant at a value of 24, leading to a constant ``effective diameter'' of $\sigma$ for two particles colliding at a 90 degree angle; Pe was then adjusted by changing $k_{\mathrm{B}}T$ (and thus $\tau_{\mathrm{r}}$). This subtle difference in how the P\'eclet number is varied may be important, especially in determining the region where phase separation arises in 3D: our preliminary results indicate that, for this particular ABP model, the shape of the phase diagram is dramatically changed when varying Pe by changing $F_{\mathrm{p}}$, possibly even removing the phase coexistence region altogether. Presumably, this effect is more visible in 3D due to the higher P\'eclet numbers required for phase separation compared to the 2D case. 

\subsection{Comparison with the continuum model}\label{Continuum_section}

\begin{figure}[h!]
  \centering
  \includegraphics[height=11cm]{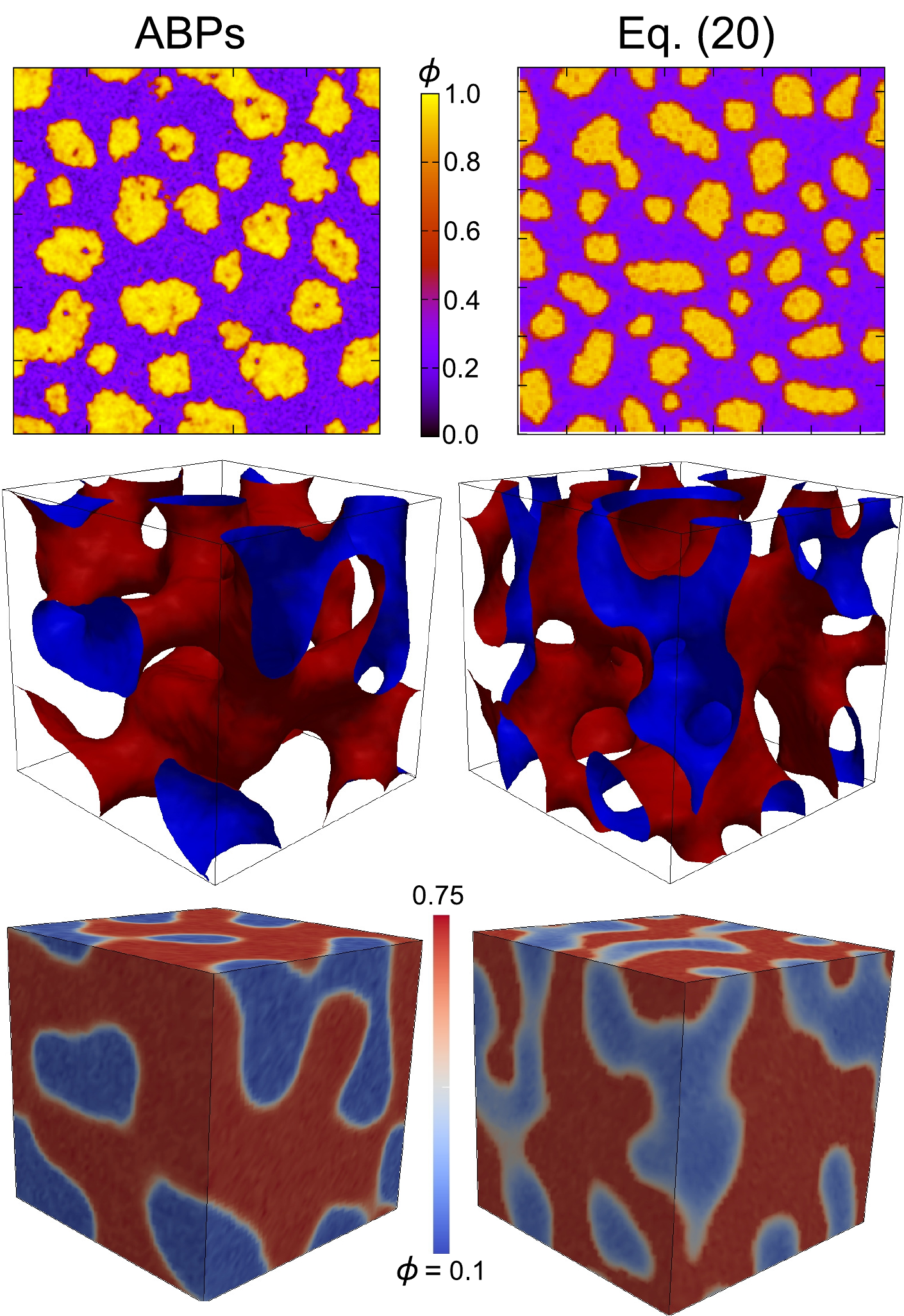}
  \caption{Snapshots obtained from an ABP simulation (left) and by numerically solving Eq. \eqref{rho_t_rescaled} (right), both at average packing fraction $\phi_0 = 0.5$. Snapshots are taken at equal times of $t = 500 \tau_{\mathrm{r}}$ ($t = 100 \tau_{\mathrm{r}}$) in 2D (3D). The continuum and ABP systems are of approximately equal size $L = 60 \lambda$ (2D, top panel) and $L \approx 21 \lambda$ (3D, center and bottom panels). The bottom panel shows 2D projections of the 3D density field, obtained from the same configurations as the isosurfaces in the center panel.}\label{Snapshots_comparison}
\end{figure}

Fig. \ref{Snapshots_comparison} shows a comparison between the results of ABP and continuum simulations at equal time and system size in 2 and 3 dimensions, at average packing fraction $\phi_0 = 0.5$. In accordance with our previous observations in 2D,\cite{Stenhammar-2013} the agreement between domain topologies observed using the ABP and continuum models is excellent, especially considering the fact that the length and time units as well as the noise level are completely determined by the mapping detailed above (with $\kappa_0$ as the only fitting parameter). In particular, the mapping successfully captures the lower apparent noise level in 3D compared to in 2D (see further movies provided as Electronic Supplementary Information$^\dag$).\footnote{A more detailed analysis of the mapping between ABPs and the continuum model shows that the lower apparent noise level in 3D compared to in 2D is simply a consequence of the fact that the number of particles $N_0$ present in a coarse-graining volume $\lambda^{d}$ is an increasing function of $d$. In other words, the ``granularity'' of the system becomes smaller in higher dimensions, leading to smaller fluctuations. Thus, this effect is not specific to non-equilibrium systems like the one studied here.} We therefore conclude that Cahn Hilliard-type continuum models like this one are successful in describing phase-ordering kinetics even in non-equilibrium systems, and at a tremendously reduced computational cost compared to direct ABP simulations: in 3D, the computational cost for solving the continuum equations for the system seen in Fig. \ref{Snapshots_comparison} is $\sim 0.5$ CPU hours, compared to the $\sim 6\times 10^{6}$ CPU hours needed for the corresponding ABP simulation. However, a striking feature observed in our 2D ABP simulations but \emph{not} captured by our continuum model is the spontaneous formation of ``gas'' voids inside the dense domains that are visible in Fig. \ref{tieline_2D}. This effect is clearly a far-from-equilibrium one in that it breaks time reversal symmetry: voids are usually formed in the center of a dense cluster and then diffuse towards the interface with the dilute phase (see further movies available as Electronic Supplementary Information$^\dag$). This intriguing effect is neither observed in our continuum simulations, nor in 3D ABP simulations, and we currently do not have a theoretical explanation for it. 

\begin{figure}[h!]
  \centering
  \includegraphics[height=10cm]{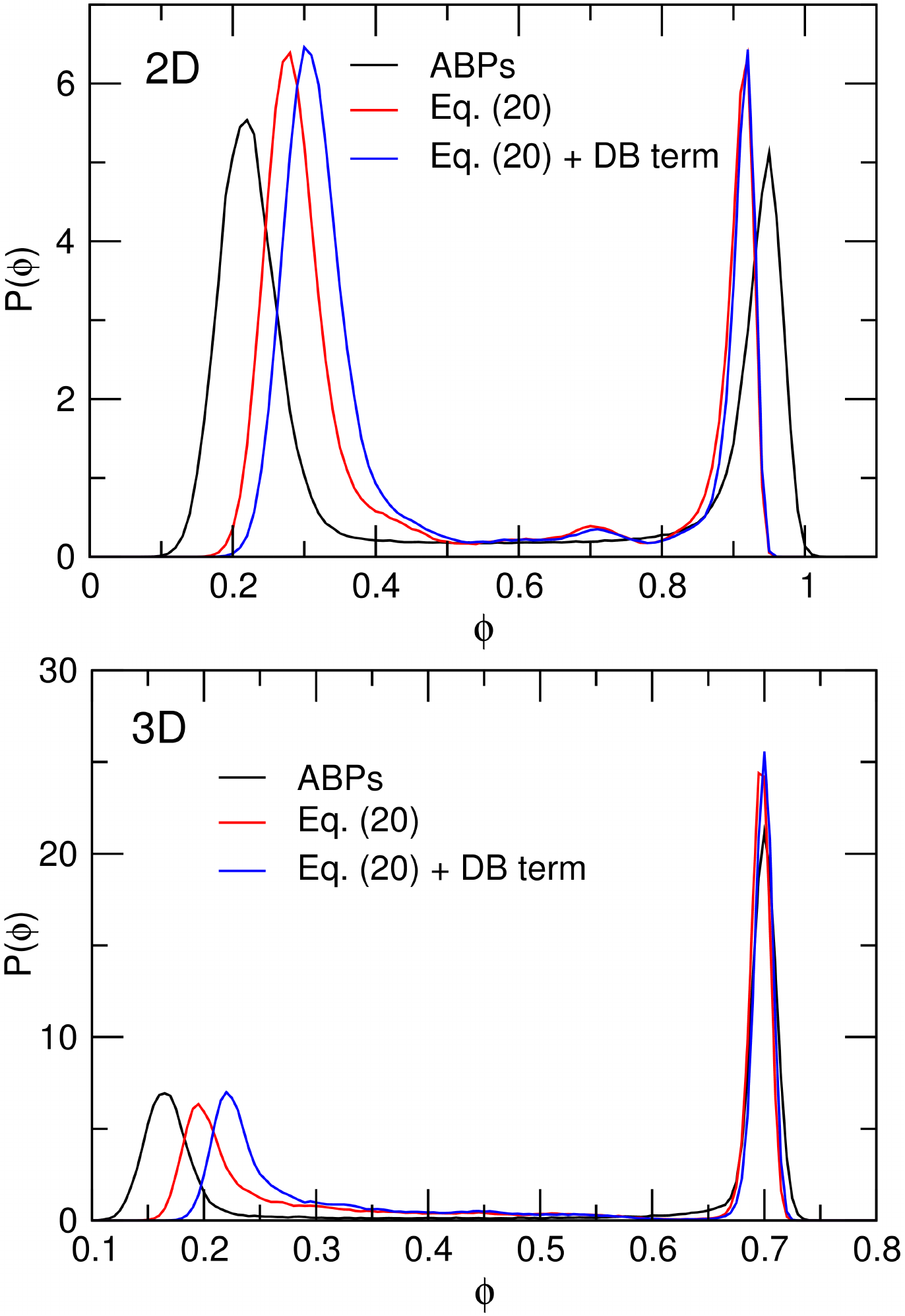} 
  \caption{Probability distribution $P(\phi)$ in 2D (top) and 3D (bottom) of the local particle packing fraction $\phi$ obtained from ABP simulations (black curves), from the continuum model as written (red curves) and with detailed balance (``DB'') restored as per Eq. \eqref{mu_db} (blue curves), all at average packing fraction $\phi_0 = 0.5$. In 2D, $P(\phi)$ was sampled over quadratic coarse-graining areas of side length $0.8\lambda$ and averaged over the time window $500\tau_{\mathrm{r}}\leq t\leq 3500\tau_{\mathrm{r}}$, while in 3D the curves were sampled from coarse-graining boxes of side length $\approx 0.4 \lambda$ at $t = 500 \tau_{\mathrm{r}}$.}\label{P_phi}
\end{figure}

\begin{figure}[h!]
  \centering
  \includegraphics[height=4.7cm]{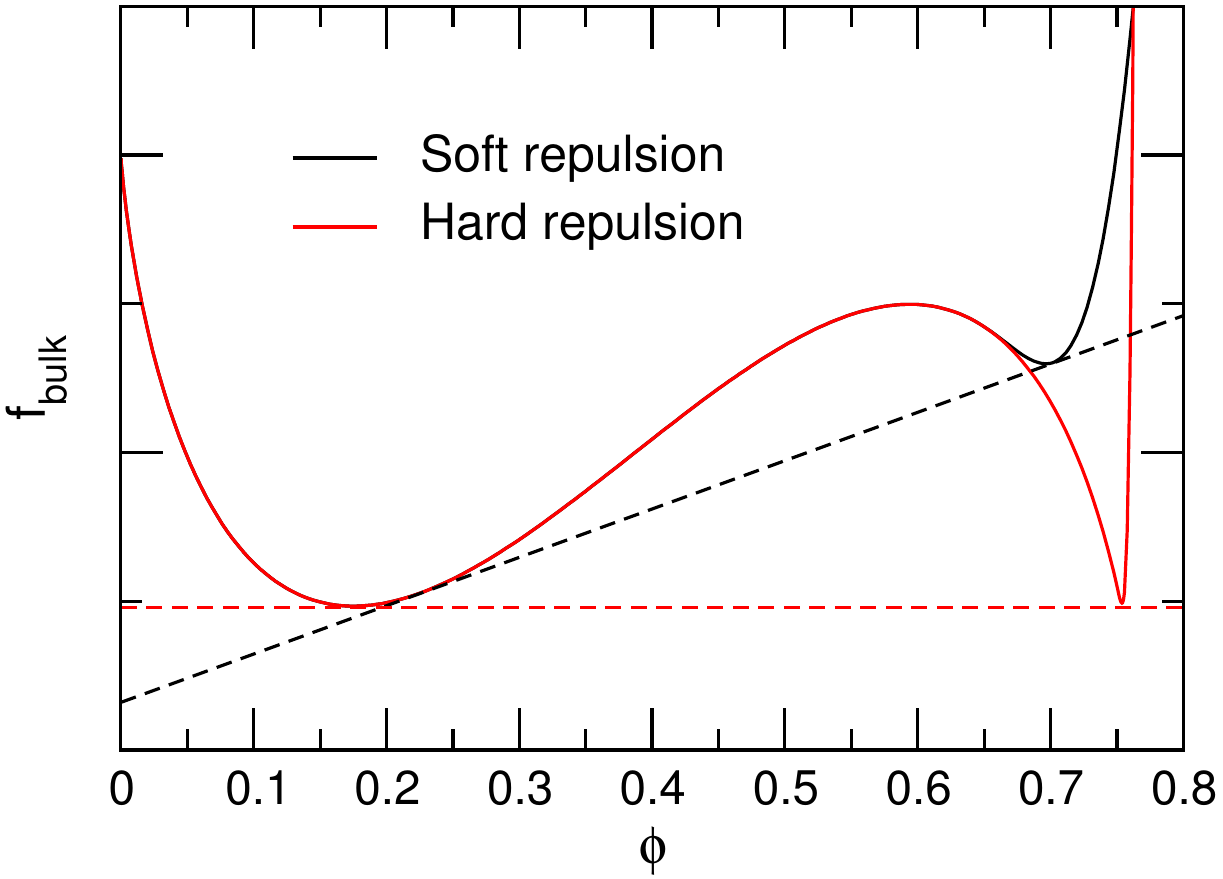}
  \caption{Bulk free energy $f_{\mathrm{bulk}} = f_0 + f_{\mathrm{rep}}$ as given by Eqs. \eqref{f_0} and \eqref{f_rep}, obtained using $v = v_0(1-1.3\phi)$, and the parameter values $\left\{ k_{\mathrm{rep}}, \phi_{\mathrm{t}} \right\} = \left\{ 10^{3}, 0.64 \right\}$ (``Soft repulsion'') and $\left\{ 10^{7}, 0.75 \right\}$ (``Hard repulsion''), respectively. The dashed lines denote approximate common-tangent constructions, yielding the coexistence densities $\left\{ \phi_{\mathrm{g}}, \phi_{\mathrm{l}} \right\} = \left\{ 0.23, 0.70 \right\}$ and $\left\{ 0.18, 0.75 \right\}$ for the ``soft'' and ``hard'' curves, respectively. Terms linear in $\phi$ are irrelevant for the common-tangent construction and have been subtracted for clarity.}\label{f_3D}
\end{figure}

Fig. \ref{P_phi} shows the probability distribution $P(\phi)$ associated with the local particle packing fraction $\phi$, at overall packing fraction $\phi_0 = 0.5$ for the continuum model and ABP simulations. In 3D as in 2D, the agreement between the general appearances of the curves is good, although the density of the dilute phase is slightly shifted to higher densities in the continuum model compared to what is observed in the ABP systems. As detailed in Fig. \ref{f_3D}, this discrepancy can be explained by the fact that the coexisting densities predicted by a common tangent construction change when we change the parameters of the \emph{ad hoc} repulsive potential of Eq. \eqref{f_rep}.\footnote{Note, however, that the common-tangent construction is only expected to yield the observed coexistence densities if the detailed balance-restoring term of Eq. \eqref{mu_db} is also added to the gradient term of the chemical potential (blue curves in Fig. \ref{P_phi}).\cite{Wittkowski-2013}} Since this potential is phenomenological, and the choice of parameters is furthermore restricted by the numerical difficulties associated with using very steep potentials, the coexistence densities (and thus the binodals and spinodals) cannot be expected to be quantitatively described by our continuum theory. Thus, the theory is not accurate when it comes to describing the location of the binodals or spinodals in the phase diagrams; for an accurate determination of phase boundaries, microscopic theories such as those developed in Refs. \citep{Redner-2013} and \citep{Bialke-2013} are better suited. 

We finally observe that the addition of the detailed balance-restoring term of Eq. \eqref{mu_db} leads to a small shift in the coexistence density of the dilute phase towards higher densities. This shift arises from the non-equilibrium nature of the system dynamics, and can be explained by theoretical arguments detailed in a recent publication.\citep{Wittkowski-2013} Its magnitude, however, turns out to be small for the present parameter set. 

\subsection{Phase separation kinetics}\label{Kinetics_section}

\begin{figure}[h!]
  \centering
  \includegraphics[height=10cm]{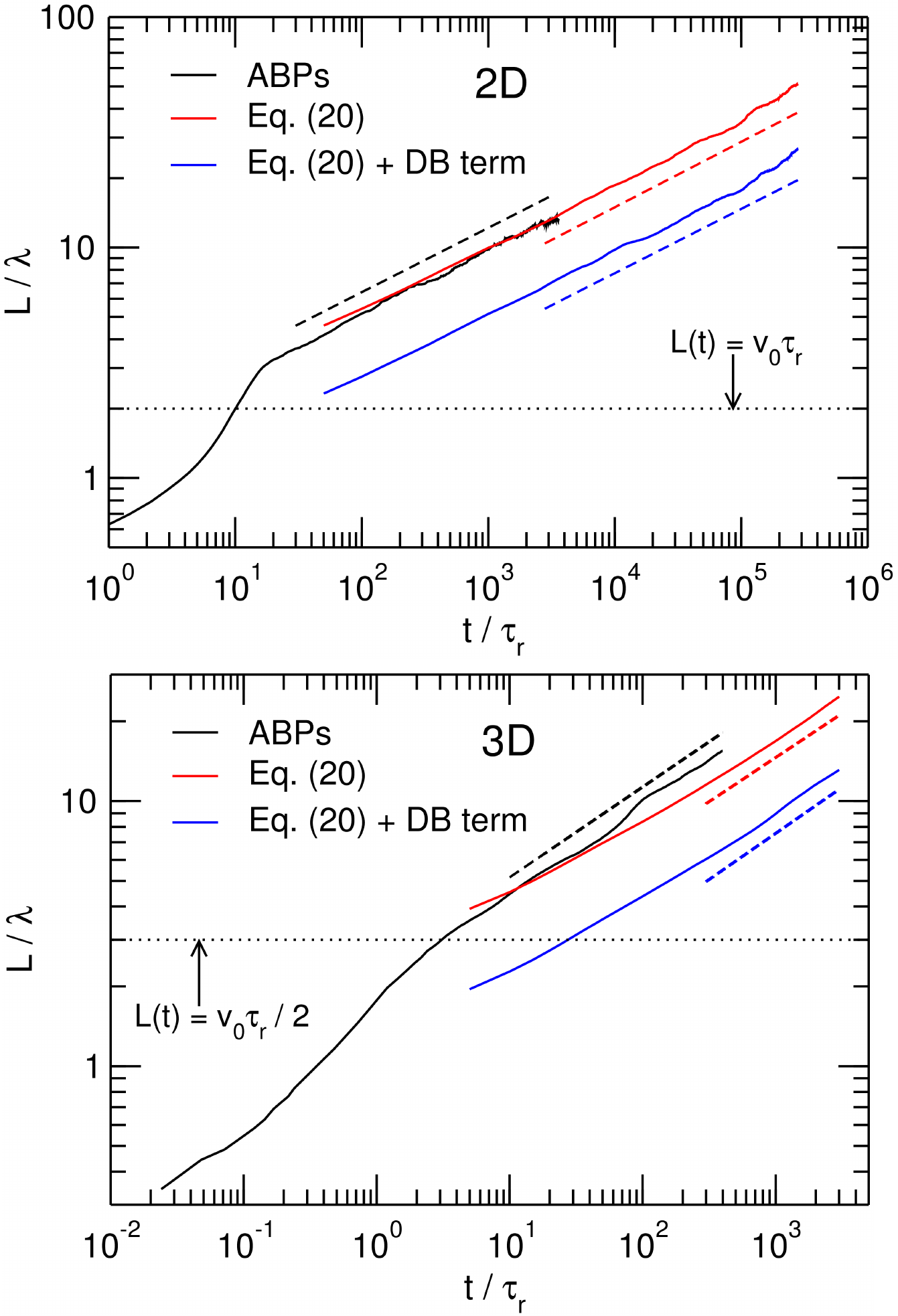}
  \caption{Time-dependent domain length $L(t)$ obtained from the inverse first moment of the structure factor at average packing fraction $\phi_0 = 0.5$ in 2D (top) and 3D (bottom). The dashed lines indicate the fitted exponents as follows: $\alpha_{\mathrm{2D}}=0.27(9)$, $\alpha_{\mathrm{3D}}=0.34(1)$ (ABPs), $\alpha_{\mathrm{2D}}=0.28(7)$, $\alpha_{\mathrm{3D}}=0.33(3)$ (continuum model), $\alpha_{\mathrm{2D}}=0.27(9)$, $\alpha_{\mathrm{3D}}=0.34(8)$ (continuum model with detailed balance term). The latter set of curves has been vertically shifted for clarity.}\label{L_t}
\end{figure}

Fig. \ref{L_t} shows the time-evolution of the characteristic domain size $L(t)$, obtained from the first moment of the static structure factor as described in Section \ref{ABP_section}, for both the continuum and ABP models. For the classical models of phase-separation kinetics, $L(t)$ usually exhibits power-law growth, \emph{i.e.}, $L(t)\sim t^{\alpha}$, where the growth exponent $\alpha$ depends on the transport properties of the system. 

For phase separation in diffusive systems with negligible hydrodynamic interactions,\footnote{Note that our simulations are distinctly different from classical molecular dynamics studies of gas-liquid coexistence in Lennard-Jones fluids (\emph{e.g.} \cite{Yamamoto-1996,Majumder-2011}), where momentum is conserved. While momentum conservation will lead to several hydrodynamic regimes with exponents $\alpha > 1/3$, our overdamped Langevin dynamics will suppress these super-diffusive transport mechanisms.} one expects $\alpha = 1/3$,\cite{Chaikin,Bray-2002,Puri} which is indeed reproduced (Fig. \ref{L_t}) within our numerical accuracy in 3D from ABP and continuum simulations. In 2D, however, both our models confirm the $\alpha \approx 0.28$ observed previously in ABP simulations by Redner \emph{et al}.\citep{Redner-2013} The fact that the 2D exponent is slightly smaller than that which is expected from traditional scaling arguments can be attributed to the relatively high noise level in the 2D system: in equilibrium systems, it is an established fact that noise will lead to a subdiffusive intermediate scaling regime, a phenomenon which is likely to be transferable to non-equilibrium phase transitions like the one studied here.\citep{Wagner-1998,Wagner-1999,Bray-2002}  We therefore conjecture that the nonstandard exponent found here is merely an intermediate regime that will eventually switch over to a $t^{1/3}$ scaling at later times. Nevertheless, it is interesting that the 3D kinetics show no evidence of this intermediate regime.
 
We also observe that the restoration of detailed balance as per Eq. \eqref{mu_db} does not seem to have any detectable effect on the phase separation kinetics for the parameters used here. However, since the detailed balance-violating term was shown in Section \ref{Continuum_section} to lead to a shift in coexistence densities, kinetic consequences of this violation must eventually arise, for some values of $\kappa_0$ or $\phi_0$. We have in fact recently found this to be the case for a continuum model containing a similar detailed balance-breaking term, where a decrease of the growth exponent was observed as the magnitude of the detailed balance violation was increased.\citep{Wittkowski-2013} Finally, it is important to note the clear crossover between superdiffusive behaviour ($\alpha > 1/3$) at short times to diffusive behaviour ($\alpha \approx 1/3$) when $L(t)$ exceeds the ``persistence length'' $\ell = v_0 \tau_{\mathrm{r}} / (d-1)$ (dotted lines in Fig. \ref{L_t}). On a microscopic level this is explained by the fact that, as long as $L < \ell$, mass transport between domains can take place ballistically, while in the region where $L > \ell$, the swimming directions of ABPs are fully randomized over the typical travelling length between domains, and thus standard diffusive behaviour is recovered. In 3D, this also highlights the problem of finite-size effects: in order to measure coarsening kinetics in the diffusive regime, the box length $L_{\mathrm{box}}$ needs to greatly exceed $\ell$, which is in turn fixed by the rather high P\'eclet number needed to obtain a deep quench in 3D ($\mathrm{Pe} = 300$ leads to $\ell = 50 \sigma$). This fact, together with the high volume fractions required for phase separation, motivates the extremely large systems ($L_{\mathrm{box}} = 350 \sigma$, $N \approx 4 \times 10^{7}$) studied here. 

\begin{figure*}[h!]
  \centering
  \includegraphics[height=11cm]{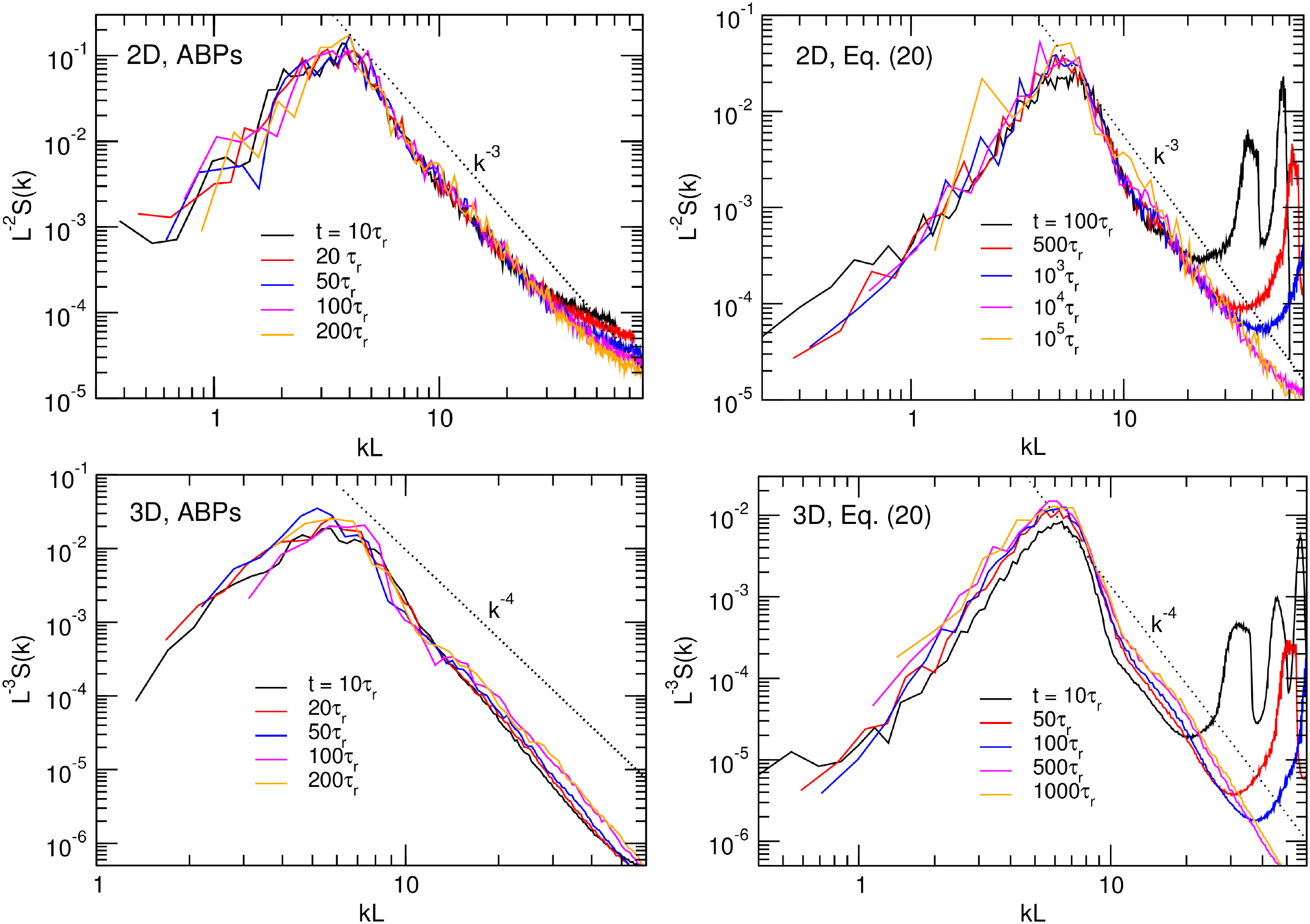}
  \caption{Static structure factor $S(k)$ obtained at different times as indicated during ABP (left) and continuum (right) simulations in 2D (top) and 3D (bottom). The curves have been rescaled in accordance with the dynamical scaling hypothesis, as described in the text. Dotted lines indicate the large-$k$ behaviour predicted by Porod's law. All results were obtained from the same systems as the results in Fig. \ref{L_t}. The oscillations at high values of k in the continuum plots correspond to very small lengthscales and are discretisation artifacts.}\label{sf}
\end{figure*}

During coarsening, we find that the static structure factor $S(k)$ typically exhibits a peak whose magnitude grows with time while its position gradually shifts towards lower values of $k$, corresponding to a growth in the typical length scale $L(t)$. In classical phase-separation theory, this behaviour is usually expressed through the so-called dynamical scaling hypothesis\cite{Chaikin,Bray-2002} as 
\begin{equation}\label{dyn_scaling}
S(k,t) = [L(t)]^{d} f(kL),
\end{equation}
where $f$ is a time-independent scaling function and $d$ is the spatial dimensionality of the system. Figure \ref{sf} shows rescaled plots of $S(k)$, sampled at different times during coarsening for both the continuum and the ABP models. The good data collapse shows that dynamical scaling is fulfilled in all cases, as has previously been observed in molecular dynamics\cite{Yamamoto-1996,Majumder-2011} as well as lattice Boltzmann\cite{Kendon-2001} simulations of equilibrium phase separations. Furthermore, the dotted lines in Fig. \ref{sf} show the prediction of Porod's law that $S(k) \sim k^{-(d+1)}$ for large $k$, as long as the interface is reasonably flat.\footnote{Strictly speaking, the structure factor should scale as $S(k) \sim k^{-(2d-\mathcal{D})}$ for large $k$, where $\mathcal{D}$ is the so-called fractal dimensionality of the interface. For a flat interface, however, $\mathcal{D} = d-1$, where $d$ is the spatial dimensionality of the system.} \citep{Bray-2002} In 2D, the correspondence is clearly satisfactory within a reasonable range of $k$ values ($10 \leq kL \leq 30$). In 3D, however, the predicted $k^{-4}$ behaviour is poorly reproduced by the ABP simulation, although it can be observed for $10 \leq kL \leq 20$ in the continuum simulation. Our explanation for this is the comparingly small spatial dimensions ($L_{\mathrm{box}} = 350 \sigma$) of the 3D ABP simulation box compared to the 2D one ($L_{\mathrm{box}} = 1000 \sigma$), which presumably does not allow the $k^{-4}$ behaviour to develop fully before boundary effects start to take over. 

\section{Conclusions and outlook}

In this study, we have performed very large scale Brownian dynamics simulations of dense suspensions containing up to 40 million active Brownian particles (ABPs), focusing on their phase behaviour and phase separation kinetics. We have also compared our results to those of a continuum model recently developed by us,\citep{Stenhammar-2013} which uses an effective free-energy mapping based on a particle swim-speed that decreases with density due to collisions. 

Our results show some remarkable similarities to phenomena seen in thermodynamic (attraction-induced) gas-liquid phase coexistence:

\begin{enumerate}
\item Both the 2D and 3D phase diagrams exhibit binodal and spinodal lines, with the role of (inverse) temperature being played by the P\'eclet number. 
\item The coexistence densities remain essentially constant along the tieline: only the amount of each phase is affected by changing the average density $\phi_0$, in accordance with what is seen in equilibrium phase diagrams.
\item The sequences of microstructures observed during coarsening closely resemble those observed in classical spinodal decompositions; this is most apparent in our largest 3D simulation (left panel in Fig. \ref{Snapshots_comparison}). 
\item The phase separation kinetics exhibit traditional power-law growth of the characteristic length-scale $L(t)$, with the 3D systems showing the classical $t^{1/3}$ behaviour. 
\item Phase separation kinetics can be accurately described by a classical continuum theory, analogous to the Cahn-Hilliard equation but using an effective free energy density that depends on $v(\phi)$. Although the effective interfacial free energy weakly violates detailed balance, the effect of this violation remains small, and only leads to a moderate shift of the coexistence densities for the parameters used here. 
\item The time-evolution of structure shows dynamical scaling behaviour. 
\end{enumerate}

The points listed above are valid in both 2 and 3 dimensions, but there are also significant differences between the two cases: 
\begin{enumerate}
\item The ``critical P\'eclet number'' needed for phase separation is significantly higher in 3D than in 2D (see Fig. \ref{phase_diagrams}). This may have practical implications for future 3D experiments, as it is not easy to experimentally control the P\'eclet number in active suspensions. 
\item The ``metastable'' region between the binodal and the spinodal is larger in 3D than in 2D; in particular, such a region does not appear to exist at all at the high-$\phi_0$ end of the 2D phase diagram.
\item There is an apparently lower level of noise in 3D than in 2D. This lower noise level in 3D leads to patterns with more well-defined geometries than in 2D (see Figs. \ref{tieline_2D} and \ref{tieline_3D}, and movies available as Electronic Supplementary Information$^\dag$). 
\item The phase transition in 3D always appears to take place via a single domain of each phase, whereas in 2D the microstructures usually consist of a continuous matrix of the majority phase containing separated domains of the minority phase which eventually coarsen through an Ostwald-like process.
\item The kinetics of domain growth differs between 2D and 3D, with the 2D case exhibiting a non-standard scaling exponent $\alpha \approx 0.28$, while in 3D the standard $\alpha \approx 1/3$ exponent is reproduced beyond the ballistic regime. 
\end{enumerate}

While our ABP model clearly relates only approximately to experimental systems, such as suspensions of bacteria or catalytic swimmers, the equilibrium-like behaviours listed above indicate surprising analogies between non-equilibrium and equilibrium phase transitions. However, several important issues within the field remain to be resolved: perhaps most crucially, the impact of hydrodynamic interactions on the complex collective behaviour of active suspensions. While the kinetic consequences of far-field hydrodynamic flows on phase-separation kinetics in classical fluids have been extensively studied, \citep{Bray-2002,Chaikin} no such studies exist for active systems (although the role of near-field hydrodynamics in the clustering of self-propelled disks has recently been investigated\citep{Zoettl-2013,Fielding-2012}). Furthermore, the subtle interplay between thermodynamic interparticle attractions and motility-induced phase transitions has only recently begun to be studied, \citep{Redner-2013-PRE} and is likely to generate many more interesting non-equilibrium phenomena. 

\section*{Acknowledgements}
Helpful discussions with Alan Bray, Aidan Brown, Tom Lion, Ignacio Pagonabarraga, Julien Tailleur, Adriano Tiribocchi, and Raphael Wittkowski are gratefully acknowledged, and JS would like to thank Fred Farrell for assistance with the numerical coarse-graining and Kevin Stratford for providing code for the calculation of structure factors. We thank EPSRC EP/J007404 for funding. JS is supported by the Swedish Research Council (350-2012-274), RJA by a Royal Society University Research Fellowship, and MEC by a Royal Society Research Professorship.

\appendix

\section{Simulation details}

\subsection{ABP model}\label{ABP_section}
Eqs.~\eqref{Langevin_t} -- \eqref{Langevin_r_3D} were solved using a slightly modified version of the LAMMPS \cite{Plimpton-1995} molecular dynamics software package. For determining the phase diagram, a cubic system with side length $150 \sigma$ (2D) or $35 \sigma$ (3D) was used, both amounting to $N \approx 20000$ particles. Periodic boundary conditions were applied in all directions. For studying the phase separation kinetics, significantly larger systems with side lengths $1000\sigma$ (2D) and $350\sigma$ (3D) were used; these system sizes amount to $N \approx 7\times 10^{5}$ and $N \approx 4\times 10^{7}$ particles in 2 and 3 dimensions, respectively. In terms of Lennard-Jones time units $\tau_{\mathrm{LJ}} = \sigma^{2}/(\varepsilon \beta D_{\mathrm{t}})$, a time step of $5 \times 10^{-5} \tau_{\mathrm{LJ}}$ was used throughout, and each simulation was run for $\sim 10^{8}$ timesteps, with the largest 3D system amounting to $\sim 700$ hours of computing time on a 8192-core IBM Blue Gene/Q node. As discussed in Section \ref{Phase_diagram_section}, the P\'eclet number was varied by varying $D_{\mathrm{t}}$ while keeping $F_{\mathrm{p}}$ constant at a value of $F_{\mathrm{p}} = 24 \varepsilon / \sigma$. Unless otherwise stated, simulations were started from an initial configuration of equilibrated passive colloids ($F_{\mathrm{p}} = 0$) with $k_{\mathrm{B}}T =\varepsilon$, and were quenched by switching on $F_\mathrm{p}$ at $t = 0$. In terms of Lennard-Jones units, the reduced length and time scales (see Section \ref{Continuum_derivation} for definitions) in the ABP systems with $\mathrm{Pe} = 100$ (2D) and $\mathrm{Pe} = 300$ (3D) are given by $\lambda_{\mathrm{2D}} = \lambda_{\mathrm{3D}} = 16.67 \sigma$ and $\tau^{(\mathrm{2D})}_{\mathrm{r}} = 1.389 \tau_{\mathrm{LJ}}$, $\tau^{(\mathrm{3D})}_{\mathrm{r}} = 4.167 \tau_{\mathrm{LJ}}$. The length scale mapping furthermore leads to $N_0 = 4 \lambda^{2} / (\pi \sigma^{2}) \approx 354$ (2D) and $N_0 = 6 \lambda^{3} / (\pi \sigma^{3}) \approx 8842$ (3D), which was used to set the noise strength when numerically solving Eq. \eqref{rho_t_rescaled}.

The characteristic domain length $L(t)$ was computed as the inverse of the first moment of the static structure factor $S(k,t)$, \emph{i.e.},
\begin{equation}
L(t) = 2\pi \left[ \frac{\int_{2\pi/L}^{k_{\mathrm{cut}}} kS(k,t) \mathrm{d}k}{\int_{2\pi/L}^{k_{\mathrm{cut}}} S(k,t) \mathrm{d}k} \right]^{-1},
\end{equation}
where $L$ is the length of the simulation box and the upper cut-off $k_{\mathrm{cut}}$ was taken to be the first minimum in $S(k)$.

In 2D, snapshots from ABP simulations were obtained by coarse-graining the local density on a grid using a weighting function $w(r) \sim \exp \left[ -r^{2}_{\mathrm{cut}}/(r^{2}_{\mathrm{cut}} - r^{2}) \right]$, where $r$ is the distance of a particle from the a particular lattice point, and $r_{\mathrm{cut}}$ is a cut-off distance which was taken to be slightly larger than the size of a lattice site. In 3D, no such weighting function was employed. The coarse-grained density field was further analyzed by constructing a density isosurface using the Paraview visualization software. 

\subsection{Continuum model}\label{Continuum_appendix}
Equations \eqref{rho_t_rescaled} -- \eqref{mu_nabla_final} were solved numerically employing a standard Euler finite-difference scheme using the parameter values reported in Table \ref{parameters}. For comparison with ABP simulations, lattice sizes of $150^{2}$ (2D) and $52^{3}$ (3D) were employed, whereas lattice sizes of $512^{2}$ (2D) and $256^{3}$ (3D) were used for the calculation growth exponents. The initial condition was taken to be one of uniform density, with a local random offset of $\approx 5\%$.  
\begin{table}[h]
\small
  \caption{Parameters used to solve Eq. \eqref{rho_t_rescaled}. $\Delta L$ indicates the lattice spacing and $\Delta t$ the size of the time step; all other parameters are defined in Section \ref{Continuum_derivation}.}
  \label{parameters}
  \begin{tabular*}{0.5\textwidth}{@{\extracolsep{\fill}}lll}
    \hline
     & 2D & 3D \\
    \hline
    $a$ & 1.0 & 1.3 \\
    $k_{\mathrm{rep}}$ & 2500 & 1000 \\
    $\phi_{\mathrm{t}}$ & 0.88 & 0.64 \\
    $\kappa_0$ & 0.3 & 0.9 \\
    $\Delta L / \lambda$ & 0.4 & 0.4 \\
    $d(d-1) \Delta t / \tau_{\mathrm{r}}$ & $1 \times 10^{-2}$ & $3 \times 10^{-2}$ \\
    \hline
  \end{tabular*}
\end{table}

\section{Density dependence of swim speed}\label{v_rho_section}

\begin{figure}[h!]
  \centering
  \includegraphics[height=10cm]{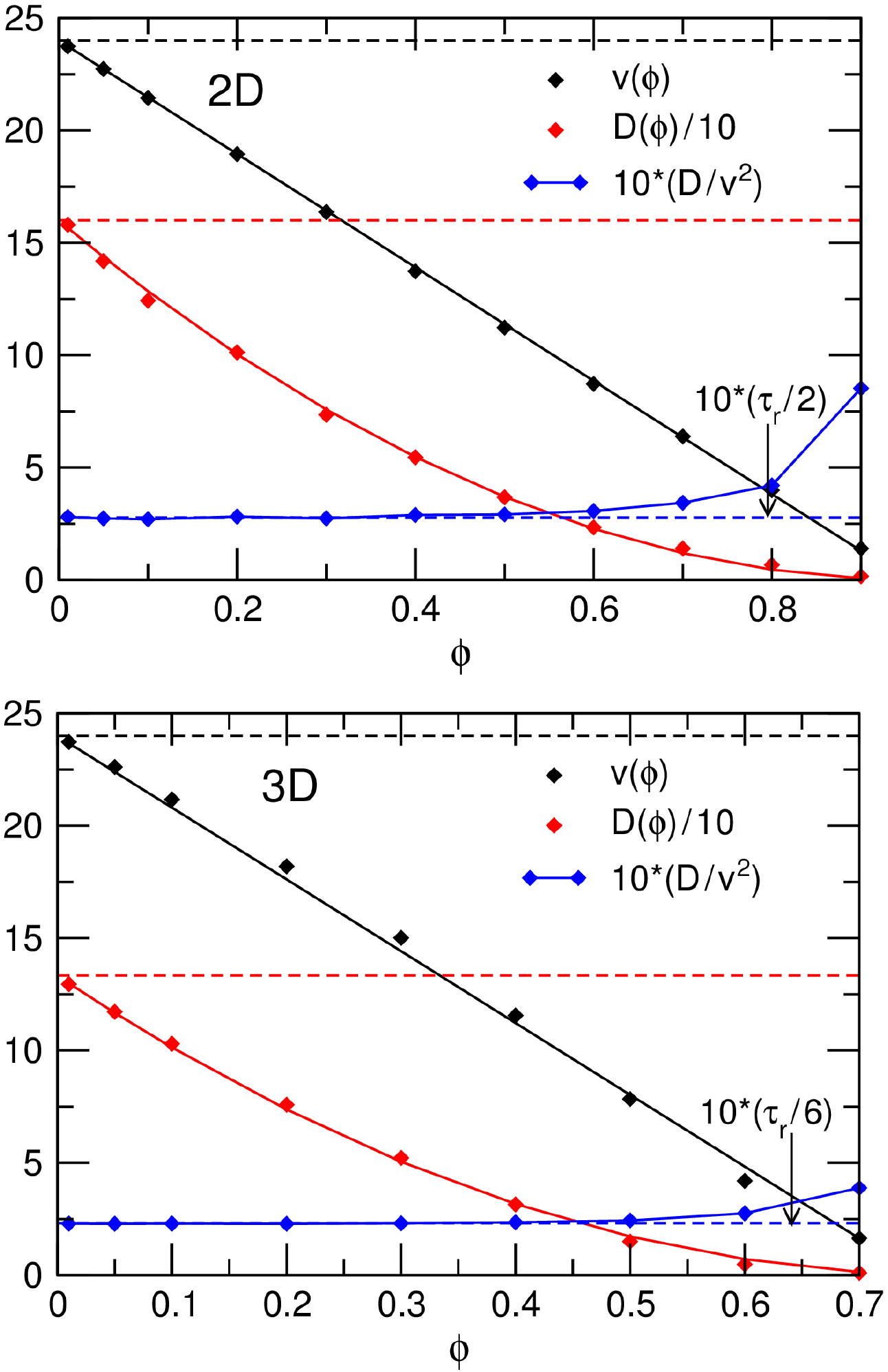}
  \caption{Density-dependent swim speed $v(\phi)$ (black symbols) and diffusivity $D(\phi)$ (red symbols), and the ratio $D/v^{2}$ (blue curve and symbols) obtained from ABP simulations at Pe = 40 (2D, top panel) and Pe = 100 (3D, bottom panel). The black and red curves show the best fits to the functions $v_0(1-a\phi)$ and $D_0(1-b\phi)^{2}$, respectively, with the optimized values $a=1.05$, $b=1.04$ (2D) and $a=1.33$, $b=1.28$ (3D). Dashed lines show the predicted zero-density values. Plotted quantities are in Lennard-Jones units $\sigma$ and $\tau_{\mathrm{LJ}}$, as defined in Section \ref{ABP_section}.}\label{v_rho_figure}
\end{figure}

As was discussed in Supplemental Information of Ref. \citep{Stenhammar-2013}, to obtain an estimate for the effective free energy, $v(\phi)$ should be sampled in the \emph{one-phase} region of the phase diagrams in Fig. \ref{phase_diagrams}, \emph{i.e.}, for sufficiently low P\'eclet numbers that the system does not phase separate. This might, in principle, lead to problems since the fitting parameters $a$ and $b$ will themselves depend on Pe. However, as we showed in Ref. \citep{Stenhammar-2013}, this dependence is small enough to be negligible. We therefore measured $v(\phi)$ and $D(\phi)$ right outside the spinodal region (for Pe = 40 and Pe = 100 in 2D and 3D, respectively). 

As our operational microscopic definition of $v(\phi)$, we sampled the average of the instantaneous velocities of all particles projected onto their self propulsion direction $\mathbf{p}_i$:
\begin{equation}
v = \beta D_{\mathrm{t}} \langle (F_{\mathrm{p}}\mathbf{p}_i + \mathbf{F}_i) \cdot \mathbf{p}_i \rangle
\end{equation}
In the limit $\phi \rightarrow 0$, where interparticle forces vanish, this definition further leads to $v \rightarrow \beta D_{\mathrm{t}} F_{\mathrm{p}} = v_0$ . The density-dependent effective diffusivity $D(\phi)$ was independently measured by linear fitting to the long-time limit of the mean-square displacement $\langle R^{2} \rangle = 2(d-1)Dt$. In Fig. \ref{v_rho_figure}, plots of $v(\phi)$ and $D(\phi)$ together with the best fits to the functions $v = v_0(1-a\phi)$ and $D = D_0(1-b\phi)^{2}$ are shown. It can clearly be seen that the predicted linear decrease of $v(\phi)$ is accurately reproduced over a wide range of packing fractions. Furthermore, the predicted relation between $D$ and $v$ is satisfactorily fulfilled, with the fitting coefficients $a$ and $b$ approximately equal in both 2D and 3D. Based on the fitted parameter values, we used $a = b = 1.0$ (2D) and $a = b = 1.3$ (3D) when solving the continuum model (see further Section \ref{Continuum_appendix}). 


\balance

%\newpage
\footnotesize{
\bibliography{bibliography} %your .bib file
\bibliographystyle{rsc} %the RSC's .bst file
}

\end{document}